\documentclass[aps,prl,reprint,superscriptaddress,nofootinbib]{revtex4-2}

\usepackage{amsmath,amssymb,bm,graphicx,mathtools,microtype,color}

\newcommand{\CN}{\mathcal C_N}
\newcommand{\LLL}{\mathrm{LLL}}

\begin{document}

\title{Spectral Crossings Diagnose Multiply Quantized Vortex Splitting}

\author{Jongheon Baek}
\thanks{Corresponding author}
\email{bjh6941@yonsei.ac.kr}
\affiliation{Department of Physics and Institute of Physics and Applied Physics, Yonsei University, Seoul 03722, Korea}

\author{Hiroyuki Tajima}
\email{hiroyuki.tajima@phys.s.u-tokyo.ac.jp}
\affiliation{Department of Physics, Graduate School of Science, The University of Tokyo, Tokyo 113-0033, Japan}
\affiliation{Quark Nuclear Science Institute, The University of Tokyo, Tokyo 113-0033, Japan}
\affiliation{RIKEN Nishina Center, Wako 351-0198, Japan}

\author{Gihwan Nam}
\email{namgh@yonsei.ac.kr}
\affiliation{Department of Physics, Yonsei University, Seoul 03722, South Korea}

\begin{abstract}
A charge-$r$ coupling vortex can remain as one fused order-$r$ zero or split into $r$ nearby unit-charge zeros after weak symmetry breaking.
The two configurations have the same total winding and the same number of dark states in the lowest-Landau-level projection, so neither the imposed orbital angular momentum nor the total winding number distinguishes their local geometry.
We study this distinction in a two-component Haldane-sphere model coupled by a Raman field.
For monopole flux $N$, the detuning $\tau_{\rm c}$ at which a defect branch crosses zero energy scales as $|\tau_{\rm c}|\sim N^{-(r+1)}$ for a fused zero and $N^{-2}$ for resolved unit zeros.
A Feshbach reduction establishes these asymptotic laws for the full multiband Hamiltonian, and numerical calculations in a monopole-harmonic basis confirm the fused-vortex scaling for $r=1,2,3$.
For a partially split cluster of diameter $d$, the scaling crosses over when $d$ is comparable to the magnetic length $\ell_B$, allowing $d$ to be inferred spectroscopically even below the spatial resolution of direct imaging.
\end{abstract}

\maketitle

Topological counting characterizes global properties of a system, but it does not generally determine the local structure.
A familiar example is provided by Landau-level physics, where the total magnetic flux fixes the number of lowest-Landau or Pauli zero modes
~\cite{AharonovCasher1979,WuYang1976,Haldane1983}, while leaving their local spatial organization undetermined.
A coupling field carrying a total winding number $r$ may realize that winding either as a single charge-$r$ singularity or as $r$ simple unit-charge singularities.
These configurations have the same total topological charge, even though their local structures are qualitatively different.


This distinction matters experimentally because total winding can remain unchanged even when the local defect structure changes.
A high-charge phase singularity is structurally fragile: weak aberrations or broken rotational symmetry can resolve a higher-order optical vortex into unit-charge singularities while conserving total winding~\cite{Shen2019,Voitiv2022}.
The analogous splitting of multiply quantized vortices has been predicted theoretically~\cite{Mottonen2003,KawaguchiOhmi2004,Kuopanportti2010} and directly observed in atomic condensates~\cite{Shin2004,Isoshima2007}.
For reviews, see Refs.~\cite{Fetter2009,KawaguchiUeda2012}.
These examples motivate a question for a coupling vortex of charge $r$: how can a fused order-$r$ zero be distinguished from $r$ nearby unit-charge zeros with the same total winding?

\begin{figure*}[t]
\centering
\includegraphics[width=0.9\textwidth]{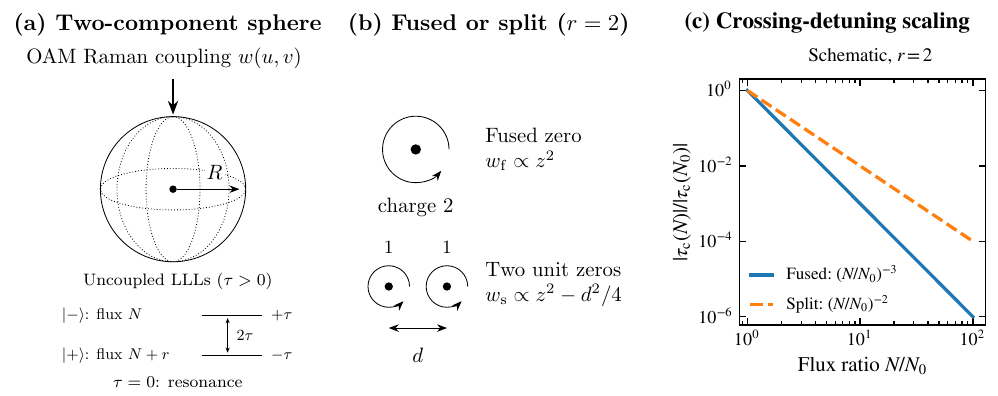}
\caption{
\textbf{Model, zero geometry, and the spectral observable.}
(a) Raman profile $w(u,v)$ couples components with fluxes $N$ and $N+r$ on a sphere of radius $R$; their LLL dimensions are $N+1$ and $N+r+1$.
The uncoupled levels $+\tau$ and $-\tau$ are shown for $\tau>0$, with separation $2\tau$ and resonance at $\tau=0$.
(b) A fused double zero and two unit zeros separated by $d$ have the same winding and LLL dark-state count.
Circular arrows indicate phase winding, not flow.
(c) Schematic $r=2$ crossing-detuning scalings, with $E(\tau_{\rm c})=0$, normalized separately at $N_0$: $(N/N_0)^{-3}$ for fused and $(N/N_0)^{-2}$ for resolved zeros.
These are asymptotic laws, not finite-$N$ data.
}
\label{fig:concept}
\end{figure*}

We address this question from a spectroscopic perspective.
Single-atom internal-state spectroscopy has reconstructed optical-field structure beyond conventional imaging resolution~\cite{Tomita2026}, motivating a spectroscopic test of defect splitting.
We consider Raman-coupled components with synthetic monopole fluxes $N$ and $N+r$ on a Haldane sphere~\cite{Haldane1983}, as shown in Fig.~\ref{fig:concept}.
Here $N$ is the integer flux of the $|-\rangle$ component, and $\tau_{\rm c}$ is the differential detuning at which a defect branch crosses zero energy.
While our argument is not limited to the sphere, such a geometry avoids physical edges; extensions beyond it are discussed in the Supplemental Material~\cite{SM}.
At fixed coupling and defect geometry, the large-$N$ crossing laws are: 
\begin{equation}
\begin{aligned}
 r~\text{resolved unit vortices}:&
 \qquad |\tau_{\rm c}|\sim N^{-2},
\\
 1~\text{fused charge-}r~\text{vortex}:&
 \qquad |\tau_{\rm c}|\sim N^{-(r+1)} .
\end{aligned}
\label{eq:punchline}
\end{equation}
The magnetic length $\ell_B\simeq\sqrt{2}\,R/\sqrt{N}$, with sphere radius $R$, sets the spatial resolution of a Landau wave packet.
A cluster of diameter $d\ll\ell_B$ appears fused, whereas resolved unit zeros are separated by more than $\ell_B$.
The slope of $\log|\tau_{\rm c}|$ versus $\log N$ changes near $N_\ast\sim(R/d)^2$.
This knee provides a spectroscopic measure of $d$, potentially below direct imaging resolution.

We establish Eq.~\eqref{eq:punchline} for resonant Landau systems and turn it into a vortex-splitting diagnostic.
Artificial gauge fields, monopole textures, and Landau quantization are now available across cold-atom and photonic platforms~\cite{Cooper2008,Dalibard2011,Goldman2014,Cooper2019,Lin2009,Aidelsburger2011,Miyake2013,Jotzu2014,Celi2014,Ray2014,Ray2015,Schine2016,Sugawa2018}.
Synthetic monopole Landau levels on a sphere and Raman transfer of optical orbital angular momentum (OAM) provide a route to this setup~\cite{Zhou2018,Andersen2006,Leanhardt2002,Chen2018,Pietila2009}.
The complete full-space construction and proof are given in the Supplemental Material \cite{SM}.

To be concrete, we consider two resonant dressed components, $\lvert-\rangle$ and $\lvert+\rangle$, that experience synthetic monopole fluxes $N$ and $N+r$, respectively.
This flux mismatch gives the two lowest Landau levels (LLLs) degeneracies $N+1$ and $N+r+1$.
Spin-dependent Berry connections provide a natural implementation because the effective monopole charge is tied to the dressed spin projection; selecting and Raman-coupling two dressed manifolds whose charges differ by $r$ realizes the required pair~\cite{Zhou2018,Pietila2009,Ray2014,Cooper2019}.
We align the uncoupled LLL energies and choose their common value as the zero of energy.
We measure energies in units of $E_0$, which for an atom of mass $m_{\rm at}$ on a sphere of radius $R$ is $E_0=\hbar^2/(8m_{\rm at}R^2)$.
In these units, the Hamiltonian is
\begin{equation}
H_{N,r}(\tau)=
\begin{pmatrix}
Q_{N}^{(-)}+\tau & \lambda M_w^\dagger
\\
\lambda M_w & Q_{N+r}^{(+)}-\tau
\end{pmatrix}.
\label{eq:H}
\end{equation}
Here $Q_N^{(-)}$ and $Q_{N+r}^{(+)}$ are the nonnegative single-component Haldane-sphere Hamiltonians with their respective LLL energies subtracted.
For flux $n=N$ or $N+r$, their eigenvalues are $4\ell(n+\ell+1)$, where $\ell=0,1,\ldots$ is the Landau-level index.
The LLL has zero energy, and the first excited level lies at $4(n+2)$.
The operator $M_w$ maps a lower-component wave function $\psi(u,v)$ to the upper component by multiplication with the spatial Raman-coupling profile $w$, so that $(M_w\psi)(u,v)=w(u,v)\psi(u,v)$.
The real parameter $\lambda$ sets the coupling strength, while the tunable detuning $\tau$ produces the opposite shifts $+\tau$ and $-\tau$ in the two components.
We denote by $\tau_{\rm c}$ the detuning at which an eigenvalue of $H_{N,r}(\tau)$ crosses zero, $E(\tau_{\rm c})=0$.

The lowest Landau orbitals can be written directly as monopole harmonics.
Using spinor coordinates $u=\cos(\theta/2)e^{-i\varphi/2}$ and $v=\sin(\theta/2)e^{i\varphi/2}$, where $\theta$ and $\varphi$ 
are the polar and azimuthal angles, respectively, a normalized flux-$N$ basis is
\begin{equation}
\begin{aligned}
\phi_m^{(N)}(u,v)
&=\left[\frac{N+1}{4\pi}\binom{N}{m}\right]^{1/2}
  u^{N-m}v^m,
\end{aligned}
\label{eq:LLLbasis}
\end{equation}
where $m=0,\ldots,N$.
The upper component has the same form with $N\to N+r$.
In a local planar patch, these states reduce to the orbitals $z^m e^{-|z|^2/(4\ell_B^2)}$, where $z=x_{\rm loc}+iy_{\rm loc}$ is the complex coordinate formed from local Cartesian coordinates and $\ell_B\simeq\sqrt{2}\,R/\sqrt{N}$.

We first show that LLL zero-mode counting depends only on $r$ and cannot distinguish fused and split configurations.
The Raman-coupling profile entering $M_w$ is the degree-$r$ spinor polynomial $w(u,v)=\sum_{j=0}^{r}w_j u^{r-j}v^j$, with complex coefficients $w_j$.
A fused vortex at one pole is
\begin{equation}
w_{\rm f}(u,v)=\mathcal N_{\rm f}v^r,
\qquad
w_{\rm f}(z)\propto z^r,
\label{eq:fused}
\end{equation}
whereas a split configuration factorizes as
\begin{equation}
w_{\rm s}(u,v)=\mathcal N_{\rm s}
\prod_{j=1}^{r}(v_j u-u_j v),
\qquad
w_{\rm s}(z)\propto\prod_{j=1}^{r}(z-z_j).
\label{eq:split}
\end{equation}
Here $\mathcal N_{\rm f}$ and $\mathcal N_{\rm s}$ are normalization factors, and $(u_j,v_j)$ are the spinor coordinates of the $j$th zero.
Both factorizations have the same total winding $r$.
Projecting the Raman-coupling operator in Eq.~\eqref{eq:H} onto the LLLs with fluxes $N$ and $N+r$, denoted by $\LLL_N$ and $\LLL_{N+r}$, gives
\begin{equation}
B_N=M_w|_{\LLL_N}:\LLL_N\longrightarrow\LLL_{N+r}.
\label{eq:BN}
\end{equation}
For nonzero \(w\), this map is injective.
Because the target manifold contains exactly \(r\) more orbitals, the
upper component has an \(r\)-dimensional dark defect space
\begin{equation}
\CN=\ker B_N^\dagger,
\qquad \dim\CN=r.
\label{eq:defect}
\end{equation}
At exact resonance, the LLL-projected coupling vanishes on $\CN$.
This LLL count depends only on $r$ and therefore cannot distinguish the two factorizations in Eqs.~\eqref{eq:fused}--\eqref{eq:split}.


Virtual transitions to higher Landau levels reveal what the LLL projection misses.
Let $P_{\CN}$ project onto $\CN$ and let $P_N^{(-)}$ project onto the lower LLL.
The reduced Landau resolvent is
\begin{equation}
G_N^{(-)}=
\left(Q_N^{(-)}|_{(\LLL_N)^\perp}\right)^{-1}
(1-P_N^{(-)}).
\label{eq:green}
\end{equation}
The restriction $Q_N^{(-)}|_{(\LLL_N)^\perp}$ acts only on states orthogonal to the lower LLL.
Its inverse therefore contains the excited Landau levels but excludes the zero-energy LLL states.
These virtual transitions generate the positive second-order operator on $\CN$,
\begin{equation}
D_N=P_{\CN}M_wG_N^{(-)}M_w^\dagger P_{\CN}.
\label{eq:D}
\end{equation}
The corresponding second-order contribution to the dark-state Hamiltonian is $-\lambda^2D_N$.
The scaling of $D_N$ follows from the spatial extent of the dark-state wave functions and the Landau-level energy denominator.
A defect state is localized over the magnetic length, $\ell_B\simeq\sqrt{2}\,R/\sqrt{N}\sim N^{-1/2}$.
Within this region around a fused order-$r$ zero, the relevant coupling matrix elements scale as $(\ell_B/R)^r\sim N^{-r/2}$ at fixed profile normalization, while the virtual Landau excitation contributes $G_N^{(-)}\sim N^{-1}$.
Hence
\begin{equation}
D_N\sim
N^{-r/2}\,N^{-1}\,N^{-r/2}=N^{-(r+1)}.
\label{eq:mechanism}
\end{equation}
At a resolved simple zero, the same argument has \(r=1\), producing the
universal \(N^{-2}\) scale for each localized branch.

For a cluster of diameter $d$ and fixed shape, write its zeros as $z_j=d\,\zeta_j$, with dimensionless $\zeta_j$ normalized to unit diameter.
The detunings $\tau_{N,a}(d)$, with branch index $a=0,\ldots,r-1$, have the leading scaling form below.
Here $\mathcal F_a(x)$ is the dimensionless crossover function, with $x=d/(\sqrt{2}\,\ell_B)\simeq d\sqrt N/(2R)$:
\begin{equation}
|\tau_{N,a}(d)|=
\lambda^2N^{-(r+1)}\mathcal F_a(x).
\label{eq:crossover}
\end{equation}
The large-$N$ limits differ: an exactly fused zero has $d=x=0$ for every $N$, whereas fixed $d>0$ gives $x\propto\sqrt N\to\infty$ and eventually resolves the unit zeros.
For $x\ll1$, $\mathcal F_a(x)$ approaches the fused constant.
For $x\gg1$, the local coupling derivatives scale as $d^{r-1}$, giving $\mathcal F_a(x)\propto x^{2(r-1)}$ and the $N^{-2}$ endpoint in Eq.~\eqref{eq:crossover}.
The $r$ branches evolve from co-located normal-jet states to one state at each simple zero.
The logarithmic slope changes between its endpoint values $r+1$ and $2$ near
\begin{equation}
N_\ast\sim\left(\frac{R}{d}\right)^2,
\qquad
\frac{d}{R}\sim N_\ast^{-1/2},
\label{eq:ruler}
\end{equation}
up to a geometry-dependent prefactor of order unity.
The crossover flux $N_\ast$ in Eq.~\eqref{eq:ruler} therefore provides a spectroscopic measure of the splitting distance.

Both asymptotic scaling laws are derived for the full multiband Hamiltonian, not merely for the LLL-projected model.
Let $\lambda_0$ and $\Lambda$ be fixed positive lower and upper bounds on the coupling strength, independent of $N$, and assume $0<\lambda_0\leq|\lambda|\leq\Lambda$.
For fixed $r$ and sufficiently large $N$, the full Hamiltonian has exactly $r$ zero-energy-crossing detunings in a window $|\tau|=O(N^{-(r+1)})$ for one fused order-$r$ zero.
Denote the positive eigenvalues of $D_N$ by $\mu_{N,0}\leq\cdots\leq\mu_{N,r-1}$.
The crossing detunings can be labeled so that
\begin{equation}
\max_{0\leq a<r}
\left|\tau_{N,a}+\lambda^2\mu_{N,a}\right|
\leq C\lambda^2N^{-(r+2)},
\label{eq:exact}
\end{equation}
where $C$ is independent of $N$ and of $\lambda$ within the stated interval.
Each defect branch is strictly decreasing in $\tau$ in this window, so its zero-crossing detuning is unique.
A nested Feshbach--Grushin reduction removes the excited Landau levels and the paired part of the LLLs without losing the $N^{-(r+1)}$ scale~\cite{SM,Feshbach,Grushin}.

\begin{figure}[t]
\centering
\includegraphics[width=\columnwidth]{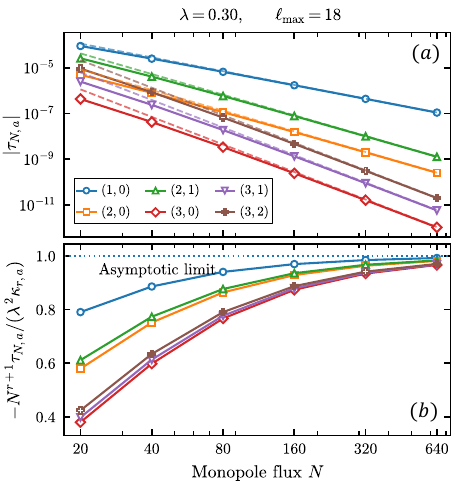}
\caption{
\textbf{Numerical validation of the fused-vortex crossing laws.}
All six branches for $r=1,2,3$ are shown at $N=20,40,80,160,320,640$, with $\lambda=0.30$ and $\ell_{\max}=18$.
The shared legend gives $(r,a)$, $a=0,\ldots,r-1$.
(a) Crossing magnitudes: symbols are numerical results, solid lines guide the eye, and dashed lines give $\lambda^2\kappa_{r,a}N^{-(r+1)}$. 
(b) Scaled crossings $-N^{r+1}\tau_{N,a}/(\lambda^2\kappa_{r,a})$ approach the dotted limit $1$.
}
\label{fig:numerics}
\end{figure}

\begin{figure*}[t]
\centering
\includegraphics[width=0.9\textwidth]{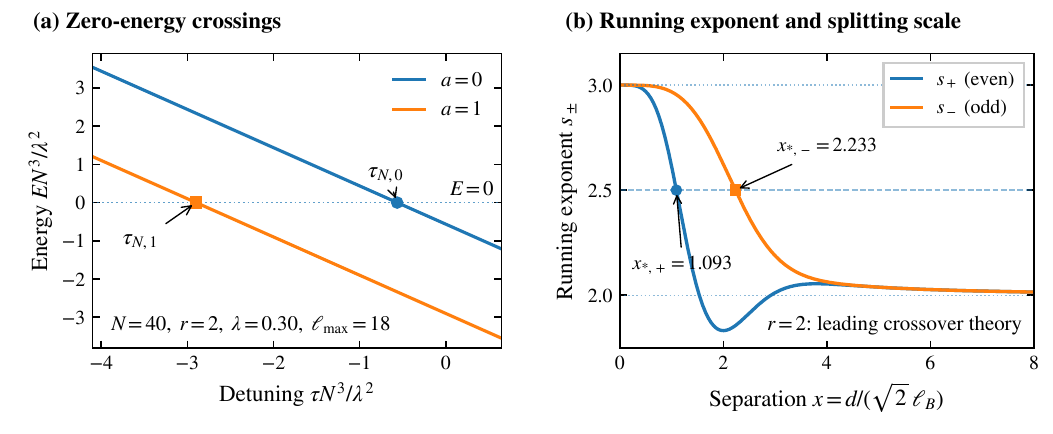}
\caption{
\textbf{Zero-energy crossings and the splitting crossover.}
(a) Fused spectrum at $N=40$, $r=2$, $\lambda=0.30$, and $\ell_{\max}=18$.
Dimensionless $E$ and $\tau$ are rescaled by $N^3/\lambda^2$; the circle and square mark $\tau_{N,0}$ and $\tau_{N,1}$ at $E=0$.
(b) Leading analytic two-zero crossover~\cite{SM}.
The even/odd parity exponents $s_\pm=-d\log|\tau_{N,\pm}|/d\log N$ are shown as functions of $x=d/(\sqrt2\ell_B)\propto\sqrt N$ with fixed $d$ and $R$, where $s_{+}$ and $s_{-}$ reduce to $s_{0}$ and $s_{1}$ at $x=0$.
Dotted lines mark the endpoint values $3$ and $2$.
The dashed $s_\pm=5/2$ line defines knees at $x_{*,+}=1.093$ and $x_{*,-}=2.233$.
The even branch briefly falls below $2$.
}
\label{fig:diagnostic}
\end{figure*}

For the fused section in Eq.~\eqref{eq:fused}, $\beta_r=\mathcal N_{\rm f}$ denotes the coefficient in $w_{\rm f}=\beta_r v^r$.
The normalization $(4\pi)^{-1}\int d\Omega\,|w_{\rm f}|^2=1$, with $d\Omega$ the solid-angle measure, gives $|\beta_r|^2=r+1$.
With this convention, the scaled crossing detunings for $0\leq a< r$ satisfy
\begin{equation}
-\frac{N^{r+1}}{\lambda^2}\tau_{N,a}
\xrightarrow{N\to\infty} \kappa_{r,a},
\qquad
\kappa_{r,a}\equiv(r+1)c_{r,a},
\label{eq:limit}
\end{equation}
where
\begin{equation}
c_{r,a}=\frac{r!}{4}
\sum_{s=0}^{a}
\binom{r}{s}\binom{a}{s}\frac{1}{r-s}.
\label{eq:c}
\end{equation}
The \(r\) coefficients are strictly ordered.
For $r=1,2,3$, the values of $\kappa_{r,a}$ are $\{0.5\}$, $\{0.75,3.75\}$, and $\{2,11,38\}$, respectively.
A fused vortex is therefore resolved twice: by the common exponent \(r+1\), and by its internal \(r\)-branch multiplet.

Monopole-harmonic calculations with cutoff $\ell_{\max}=18$ test Eq.~\eqref{eq:limit} for all six fused branches with $r=1,2,3$ [Fig.~\ref{fig:numerics}].
The $r=1$ case supplies the simple-zero endpoint.
At $N=640$ and $\lambda=0.30$, representative ratios to the asymptotic coefficients are $0.992$, $0.983$, and $0.970$ for $r=1,2,3$, respectively.
At $N=160$, changing $\lambda$ from $0.15$ to $0.30$ changes the scaled roots by less than $4.0\times10^{-9}$ relatively in all channels, consistent with leading quadratic coupling dependence.

To estimate the physical signal, we restore the energy unit $E_0$ defined above.
The first Landau gap is $\Delta_{\rm LL}=4E_0(N+2)\simeq4NE_0$, and the physical crossing shift is $E_0|\tau_{N,a}|$.
Equation~\eqref{eq:limit} then gives
\begin{equation}
\frac{E_0|\tau_{N,a}|}{\Delta_{\rm LL}}
\simeq
\frac{\lambda^2\kappa_{r,a}}{4N^{r+2}}.
\label{eq:units}
\end{equation}
For $\Delta_{\rm LL}/h=1~\mathrm{kHz}$ and $\lambda=0.30$, a fused $r=2$ zero at $N=20$ gives $E_0|\tau|/h\simeq0.11$ and $0.53~\mathrm{mHz}$; $r=3$ is substantially more demanding.
Optical-lattice Sr clocks have demonstrated submillihertz frequency precision~\cite{Marti2018}, although resolving these crossings remains challenging.
At the photonic Landau gap $\omega_c/2\pi=2.1671~\mathrm{GHz}$ of Ref.~\cite{Schine2016}, the same ratios give $0.23$ and $1.14~\mathrm{kHz}$.
The reported $200~\mathrm{kHz}$ cavity linewidth still requires a narrower or phase-sensitive realization.

A related implementation may also be possible in a planar Landau-level geometry, provided that the relevant bulk states remain well separated from higher Landau levels and from physical-edge effects.
Such a realization would require Raman coupling with the appropriate local zero structure.
Raman-induced spin--orbital-angular-momentum coupling and vortex textures have already been realized in spinor Bose--Einstein condensates~\cite{Chen2018}, providing experimental ingredients that may be useful for realizing related planar configurations.

The diagnostic uses the slope of $|\tau_{N,a}|$ rather than its fully converged coefficient.
For $r=1$, $a=0$, and $N=40$, the normalized crossing in Fig.~\ref{fig:numerics} is $0.886$, while the unscaled curve approaches the predicted power.
Fitting moderate fluxes avoids the rapid signal suppression in Eq.~\eqref{eq:units}, but must account for finite-$N$ corrections.

The measurement protocol is to determine the crossing detunings at several fluxes.
For each $N$, one may sweep $\tau$ and locate the zero-energy crossing of each defect branch.
Figure~\ref{fig:diagnostic}(a) shows the two defect branches and their zero crossings for a fused $r=2$ zero at $N=40$, calculated with the same $\ell_{\max}=18$ cutoff as Fig.~\ref{fig:numerics}.
The resulting logarithmic slope is
\begin{equation}
s_a\xrightarrow{N\to\infty}
\begin{cases}
r+1, & d=0,\\
2, & d>0\ \text{fixed},
\end{cases}
\quad
s_a\equiv-\frac{d\log|\tau_{N,a}|}{d\log N}.
\label{eq:inverse}
\end{equation}
Changing $\lambda$ rescales the crossing amplitudes by $\lambda^2$ at leading order, not their asymptotic exponents.
For fixed $d>0$, increasing $N$ resolves the zeros and yields $d/R\sim N_\ast^{-1/2}$ up to the geometry-dependent prefactor in Eq.~\eqref{eq:ruler}.
Figure~\ref{fig:diagnostic}(b) shows the branch-dependent knees and a nonmonotonic crossover of even/odd parity exponents $s_{\pm}$ at $r=2$.
Identifying the endpoint powers needs no absolute calibration of $\lambda$ or $\beta_r$.
The spectral method compares $d$ with $\ell_B$; state-resolved imaging locates the zeros when spatial resolution permits~\cite{SM}.

We note that the present vortex-splitting diagnostic relies on resolving the resonance spectroscopically and is not protected against arbitrary component-dependent disorder.
A component-dependent diagonal perturbation of order unity or $N^{-1}$ can mask the $N^{-(r+1)}$ hierarchy and must be removed by matching the two resonant manifolds.
Once this background is calibrated, the exponent is fixed by the local order of the coupling zero.
The spherical geometry is not essential under the magnetic-gap, Bergman-kernel, and holomorphic-defect hypotheses stated in the Supplemental Material~\cite{SM}.
Under these assumptions, an order-$r$ coupling zero generates an $N^{-(r+1)}$ crossing cluster, while the total branch count retains only its integrated multiplicity~\cite{Finski2024,MaMarinescu2012}.

In summary, we have proposed a way to resolve local vortex structures beyond the resolution of real-space imaging. 
A fused charge-$r$ zero and $r$ resolved unit zeros have the same total winding and LLL dark-state count, but distinct full-spectrum crossing laws: $N^{-(r+1)}$ and $N^{-2}$.
The leading crossover provides a spectroscopic measure of their separation, potentially below direct imaging resolution.

\vspace{5pt}
\textit{Acknowledgments}--- H.~T. was supported by Grants-in-Aid for Scientific Research provided by JSPS through Grants Nos.~JP22K13981, JP23K22429, and JP26K07063. J.~B. was supported by the National Research Foundation of Korea (NRF) under Grant No. RS-2023-NR077232.

\end{document}


\title{Supplemental Material for\\
``Spectral Crossings Diagnose Multiply Quantized Vortex Splitting''}

\author{Jongheon Baek}
\thanks{Corresponding author}
\email{bjh6941@yonsei.ac.kr}
\affiliation{Department of Physics and Institute of Physics and Applied Physics, Yonsei University, Seoul 03722, Korea}

\author{Hiroyuki Tajima}
\email{hiroyuki.tajima@phys.s.u-tokyo.ac.jp}
\affiliation{Department of Physics, Graduate School of Science, The University of Tokyo, Tokyo 113-0033, Japan}
\affiliation{Quark Nuclear Science Institute, The University of Tokyo, Tokyo 113-0033, Japan}
\affiliation{RIKEN Nishina Center, Wako 351-0198, Japan}

\author{Gihwan Nam}
\email{namgh@yonsei.ac.kr}
\affiliation{Department of Physics, Yonsei University, Seoul 03722, South Korea}

\maketitle

This Supplemental Material gives the model definitions, derivations, and numerical procedures used in the main text.
Section~\ref{sec:model} defines the two-component Haldane-sphere model and identifies the lowest-Landau-level (LLL) dark-state space.
Sections~\ref{sec:fused}--\ref{sec:resolved} derive the second-order effective operator and the full-spectrum crossing laws for fused and resolved zeros.
Section~\ref{sec:crossover} derives the leading splitting crossover and its analytic solution for two symmetric zeros.
Sections~\ref{sec:inverse}--\ref{sec:data} explain the spectroscopic protocol, physical units, assumptions beyond the sphere, and numerical data.

We use ``coupling vortex'' for a zero of the complex Raman-coupling profile together with its phase winding.
A fused order-$r$ zero has multiplicity $r$ at one point, whereas a resolved unit-charge zero is a simple zero separated from the others on the magnetic-length scale.
The terms ``dark state'' and ``defect state'' refer to the same upper-LLL subspace defined below; they are dark in the LLL-projected coupling, but acquire energy shifts through higher Landau levels.
A crossing detuning is a value of the tunable differential detuning $\tau$ at which a Hamiltonian eigenvalue passes through zero, not an avoided-crossing gap.

\section{Haldane-sphere model and LLL dark states}
\label{sec:model}

\subsection{Monopole harmonics, energy units, and magnetic length}

Let the sphere have radius $R$, and let one internal component experience an integer synthetic monopole flux $n\geq0$.
Here $n$ denotes a generic flux; the two components used below have $n=N$ and $n=N+r$.
The polar and azimuthal angles $\theta$ and $\varphi$ define the spinor coordinates
\begin{equation}
 u=\cos\frac{\theta}{2}\,e^{-i\varphi/2},\qquad
 v=\sin\frac{\theta}{2}\,e^{i\varphi/2}.
 \label{eq:spinor}
\end{equation}
The lowest Landau level (LLL) has degeneracy $n+1$.
We denote this space by $\LLL_n$ and use the basis
\begin{equation}
 \phi_m^{(n)}(u,v)=
 \left[\frac{n+1}{4\pi}\binom{n}{m}\right]^{1/2}
 u^{n-m}v^m,
 \qquad m=0,\ldots,n,
 \label{eq:LLLbasisS}
\end{equation}
normalized with the solid-angle measure $d\Omega$.
The full Landau basis can also be written as normalized Wigner functions $|j,m,n/2\rangle=\sqrt{2j+1}\,D^j_{m,n/2}$, with $j=n/2+\ell$ and Landau-level index $\ell=0,1,\ldots$~\cite{WuYang,Haldane,Varshalovich}.
These Wigner states use the normalized Haar-measure convention, and their magnetic label $m$ is distinct from the polynomial orbital index in Eq.~\eqref{eq:LLLbasisS}.

As in the main text, we subtract the uncoupled LLL energies, align them at resonance, and measure Hamiltonian eigenvalues and detunings in units of $E_0$.
For an atom of mass $m_{\rm at}$ on the sphere, $E_0=\hbar^2/(8m_{\rm at}R^2)$.
The dimensionless Landau energies are
\begin{equation}
 E_{n,\ell}=4\ell(n+\ell+1).
 \label{eq:LandauEnergy}
\end{equation}
The physical energy is $E_0E_{n,\ell}$, and the physical gap above the LLL is
\begin{equation}
 \Delta_{\rm LL}(n)=4E_0(n+2)=4nE_0[1+\Ord(n^{-1})].
 \label{eq:gap}
\end{equation}
Thus $\Delta_{\rm LL}\simeq4NE_0$ for the flux-$N$ component $|-\rangle$ at large $N$.
Unless a factor $E_0$ is written explicitly, all Hamiltonians, resolvents, energies, and detunings below use the dimensionless convention.

Near the north pole, let $z=x_{\rm loc}+iy_{\rm loc}$ be a physical local Cartesian coordinate, with $z\simeq R\theta e^{i\varphi}$ at leading order.
After removing the local gauge phase, Eq.~\eqref{eq:LLLbasisS} approaches the planar orbitals $z^m\exp[-|z|^2/(4\ell_B^2)]$.
The magnitude $|u|^N=\exp[-N\theta^2/8+\Ord(N\theta^4)]$ gives
\begin{equation}
 \ell_B=\frac{\sqrt2 R}{\sqrt N}[1+\Ord(N^{-1})].
 \label{eq:lB}
\end{equation}
The local coordinate $z$ has dimensions of length; the rescaled coordinate $Z=z/(\sqrt2\ell_B)$ introduced below is dimensionless.

The LLL admits a reproducing kernel: a state centered at a point can be used to evaluate any LLL wave function at that point.
In a common local gauge, the kernel is
\begin{equation}
 K_n(p,q)=\frac{n+1}{4\pi}
 \bigl(\bar u_pu_q+\bar v_pv_q\bigr)^n,
 \label{eq:Bergmankernel}
\end{equation}
where $p$ and $q$ are points on the sphere, and $(u_p,v_p)$ and $(u_q,v_q)$ are their spinor coordinates.
Write $K_p^{(n)}(q)=K_n(p,q)$ and $\widehat K_p^{(n)}=K_p^{(n)}/\|K_p^{(n)}\|$.
The reproducing property is $\langle s,K_p^{(n)}\rangle=s(p)$ for an LLL state $s$.
The normalized states $\widehat K_p^{(n)}$ are the coherent states used below.
For two points at geodesic distance $d(p,q)$, their overlap satisfies
\begin{equation}
 \left|\left\langle\widehat K_p^{(n)},\widehat K_q^{(n)}\right\rangle\right|
 =\cos^n\!\left(\frac{d(p,q)}{2R}\right).
 \label{eq:kerneloverlap-general}
\end{equation}
This identity gives a spatial width of order $\ell_B$.
For distinct points whose separation is fixed as $N$ grows, the overlap decreases exponentially.

\subsection{Spin-dependent fluxes and the Raman-coupling profile}

We retain two dressed internal components $| -\rangle$ and $|+\rangle$ with fluxes $N$ and $N+r$.
The terms ``lower and ``upper below label the flux-$N$ and flux-$(N+r)$ components, respectively, not a fixed ordering of their energies.
A dressed state $|\chi_\sigma(\bm r)\rangle$, with component label $\sigma\in\{-,+\}$, depends on position and carries the Berry connection
\begin{equation}
 \bm A_\sigma=i\langle\chi_\sigma|\bm\nabla\chi_\sigma\rangle,
 \qquad
 \frac{1}{2\pi}\int_{S^2}d\bm A_\sigma=N_\sigma.
 \label{eq:Berry}
\end{equation}
Here $d\bm A_\sigma$ denotes the Berry-curvature two-form, and $N_\sigma$ is its integrated flux.
Selecting $N_-=N$ and $N_+=N+r$ gives the required mismatch; Raman-dressed spin textures provide a route to such spin-dependent connections~\cite{Pietila,Zhou,Ray}.
The Raman profile must transform between the two flux sectors.
Mathematically, it is a section of their difference line bundle; on the sphere, the holomorphic profile used here is a degree-$r$ spinor polynomial,
\begin{equation}
 w(u,v)=\sum_{k=0}^{r}w_k u^{r-k}v^k,
 \label{eq:wpoly}
\end{equation}
where the coefficients $w_k$ are complex.
This is the spatial coupling profile, rather than the wave function of an atomic condensate.

With the common uncoupled LLL energy set to zero, the dimensionless Hamiltonian is
\begin{equation}
 H_{N,r}(\tau)=
 \begin{pmatrix}
 Q_N^{(-)}+\tau & \lambda M_w^\dagger\\
 \lambda M_w & Q_{N+r}^{(+)}-\tau
 \end{pmatrix}.
 \label{eq:HS}
\end{equation}
The nonnegative single-component operators $Q_N^{(-)}$ and $Q_{N+r}^{(+)}$ have the spectra in Eq.~\eqref{eq:LandauEnergy} with $n=N$ and $n=N+r$, respectively.
The multiplication operator acts as $(M_w\psi)(u,v)=w(u,v)\psi(u,v)$, mapping the $|-\rangle$ component to the $|+\rangle$ component.
The real parameter $\lambda$ sets the coupling strength, and $\tau$ shifts the $|-\rangle$ and $|+\rangle$ components by $+\tau$ and $-\tau$, respectively.
Their uncoupled LLL energies are $+\tau$ and $-\tau$, respectively, with resonance at $\tau=0$; the flux-$N$ level is higher for $\tau>0$, as illustrated in the main text.
We label the $r$ defect branches by $a=0,\ldots,r-1$ and write $\tau_{N,a}$ for the detuning at which branch $a$ crosses zero energy, $E_a(\tau_{N,a})=0$; $\tau_{\rm c}$ denotes a crossing without specifying its branch.
The physical energy and detuning scales are obtained by multiplication by $E_0$.
The term ``full multiband Hamiltonian'' refers to Eq.~\eqref{eq:HS} with all Landau levels retained.
Numerical matrices with a finite cutoff $\ell_{\max}$ are distinguished from this full operator.

The fused and split profiles are
\begin{equation}
 w_{\rm f}(u,v)=\mathcal N_{\rm f}v^r,
 \qquad
 w_{\rm s}(u,v)=\mathcal N_{\rm s}
 \prod_{j=1}^{r}(v_j u-u_jv).
 \label{eq:fused-split-S}
\end{equation}
Here $\mathcal N_{\rm f}$ and $\mathcal N_{\rm s}$ fix the profile normalization, and $(u_j,v_j)$ are the spinor coordinates of the $j$th zero.
The first profile has one order-$r$ zero; the second has $r$ simple zeros when the zero positions are distinct.
Both profiles have total degree and winding $r$.

\subsection{Projected coupling, dark-state count, and local basis}

Because $w$ is holomorphic, multiplication maps the lower LLL exactly into the upper LLL:
\begin{equation}
 B_N=M_w|_{\LLL_N}:\LLL_N\longrightarrow\LLL_{N+r}.
 \label{eq:BN-S}
\end{equation}
For nonzero $w$, this map is injective: if $ws$ is the zero polynomial, the lower-LLL polynomial $s$ must vanish identically.
The upper LLL contains $r$ more orbitals than the lower LLL.
Its subspace orthogonal to the image of $B_N$ is therefore
\begin{equation}
 \CN=\ker B_N^\dagger,
 \qquad
 \dim\CN=(N+r+1)-(N+1)=r.
 \label{eq:defect-S}
\end{equation}
States in $\CN$ are not coupled to the lower LLL by $B_N^\dagger$ and are the LLL dark states.
This count depends only on total degree, not on whether the zeros are fused or separated.
In either configuration, the bare LLL-projected dark levels have energy $-\tau$ and cross zero at $\tau=0$; the fine crossing shifts arise from the excited-level processes described in Sec.~\ref{sec:fused}.

Their spatial structure is described by the coherent states defined above.
If $p_j$ is a simple zero of $w$, the reproducing property gives
\begin{equation}
 \langle B_Ns,K_{p_j}^{(N+r)}\rangle=(ws)(p_j)=0
 \label{eq:kernelzero-general}
\end{equation}
for every lower-LLL state $s$.
Thus $K_{p_j}^{(N+r)}$ belongs to $\CN$.
For $r$ distinct zeros, these states are linearly independent and span the full dark-state space:
\begin{equation}
 \CN=\operatorname{span}\{K_{p_1}^{(N+r)},\ldots,K_{p_r}^{(N+r)}\}.
 \label{eq:simpleframe}
\end{equation}
They form a basis, but need not be orthogonal at finite separation.

For a zero of multiplicity $m$ at $p$, value evaluation is supplemented by derivative evaluation.
Define the derivative, or jet, states $J_{p,a}^{(N+r)}$ by
\begin{equation}
 \langle s,J_{p,a}^{(N+r)}\rangle=(\nabla_z^a s)(p),
 \qquad a=0,\ldots,m-1,
 \label{eq:jetkernel}
\end{equation}
where $s$ is an upper-LLL state and $\nabla_z$ is its covariant derivative in the local complex coordinate.
The index $a$ specifies the derivative order, including value evaluation at $a=0$.
For every lower-LLL polynomial $t$, the product $wt$ vanishes to order $m$ at $p$, so its derivatives of orders $0,\ldots,m-1$ vanish there.
The corresponding jet states therefore lie in $\CN$.
Counting all zeros with their multiplicities gives $r$ states spanning $\CN$.
As distinct zeros merge, suitable combinations of their coherent states approach these derivative states; this is the confluent-frame limit.
For one fused zero, the $r$ normal-jet branches are localized at the same point rather than at $r$ separate centers.

For $w_{\rm f}=\mathcal N_{\rm f}v^r$, Eq.~\eqref{eq:LLLbasisS} gives the exact
action
\begin{equation}
 B_N\phi_m^{(N)}=s_{N,m}\phi_{m+r}^{(N+r)},
 \qquad
 s_{N,m}=\mathcal N_{\rm f}
 \left[\frac{N+1}{N+r+1}
 \frac{\binom Nm}{\binom{N+r}{m+r}}\right]^{1/2}.
 \label{eq:exactsingular}
\end{equation}
Consequently
\begin{equation}
 \CN=\operatorname{span}\{\phi_0^{(N+r)},\ldots,\phi_{r-1}^{(N+r)}\},
 \label{eq:fusedbasis}
\end{equation}
and
\begin{equation}
 |s_{N,m}|^2=|\mathcal N_{\rm f}|^2\frac{N+1}{N+r+1}
 \frac{(m+1)\cdots(m+r)}{(N+1)\cdots(N+r)}.
 \label{eq:singularproduct}
\end{equation}
The singular values of $B_N$ are $|s_{N,m}|$; the phase of $\mathcal N_{\rm f}$ does not affect them.
For fixed $r$ and fixed nonzero profile normalization, there are positive constants $c_r,C_r$, independent of $N$, such that
\begin{equation}
 c_rN^{-r}\one\le B_N^\dagger B_N\le C_r\one.
 \label{eq:vanishingcomparison-explicit}
\end{equation}
Here $\one$ is the identity on $\LLL_N$.
This bound says that the smallest squared coupling amplitude scales as $N^{-r}$, while the largest remains bounded.
It is the vanishing-symbol estimate needed when the paired LLL states are eliminated in Sec.~\ref{sec:fullcrossing}.

\section{Second-order effective operator and fused-branch coefficients}
\label{sec:fused}

\subsection{Virtual transitions outside the LLL}

Let $P_N^{(-)}$ project onto the lower LLL and let $P_{\CN}$ project onto the upper-component dark-state space $\CN$.
The reduced lower-component Landau resolvent is
\begin{equation}
 G_N^{(-)}=
 \left(Q_N^{(-)}|_{(\LLL_N)^\perp}\right)^{-1}
 (1-P_N^{(-)}).
 \label{eq:Green-S}
\end{equation}
The inverse is taken only on states orthogonal to the lower LLL, where the dimensionless Landau energies are positive.
It is extended by zero on the LLL, so the zero-energy states do not enter its denominators.
Virtual transitions to these excited states generate the positive second-order effective operator
\begin{equation}
 D_N=P_{\CN}M_wG_N^{(-)}M_w^\dagger P_{\CN}\geq0.
 \label{eq:DN-S}
\end{equation}
Read from right to left, the operators take a dark state into the lower excited levels, weight each contribution by its inverse energy, and return it to $\CN$.
The corresponding contribution to the effective dark-state Hamiltonian is $-\lambda^2D_N$, not $+\lambda^2D_N$.
Positivity follows from the factorization
\begin{equation}
 D_N=
 \left[(G_N^{(-)})^{1/2}M_w^\dagger P_{\CN}\right]^\dagger
 \left[(G_N^{(-)})^{1/2}M_w^\dagger P_{\CN}\right].
 \label{eq:DNfactor}
\end{equation}
For a nonzero degree-$r$ profile, $D_N$ is strictly positive on $\CN$.
If $D_Nu=0$, the factorization gives $(1-P_N^{(-)})M_w^\dagger u=0$.
The remaining LLL component also vanishes because $P_N^{(-)}M_w^\dagger u=B_N^\dagger u=0$ for $u\in\CN$.
Thus $M_w^\dagger u=\overline w\,u=0$ almost everywhere.
Since a nonzero holomorphic profile has only finitely many zeros on the sphere, this requires $u=0$.

A defect state is localized over a distance of order $\ell_B$.
At fixed $R$ and fixed profile normalization, an order-$r$ zero gives a coupling amplitude of order $(\ell_B/R)^r\sim N^{-r/2}$.
The two coupling factors in Eq.~\eqref{eq:DN-S} therefore give $N^{-r}$, and the Landau resolvent supplies another $N^{-1}$.
The resulting scale is $D_N=\Ord(N^{-r-1})$.
The local oscillator calculation below determines its branch-dependent coefficients.

\subsection{Local Bargmann--Fock limit}

The magnetic-length rescaling $Z=z/(\sqrt2\ell_B)$ gives a planar oscillator description near the zero.
This is the Bargmann--Fock limit: the spherical LLL becomes the planar LLL, while the excited Landau levels become cyclotron oscillator states.
Introduce commuting guiding-center and cyclotron annihilation operators $\mathfrak a$ and $\mathfrak b$:
\begin{equation}
 [\mathfrak a,\mathfrak a^\dagger]=
 [\mathfrak b,\mathfrak b^\dagger]=1,
 \qquad
 [\mathfrak a,\mathfrak b]=[\mathfrak a,\mathfrak b^\dagger]=0.
 \label{eq:oscillators}
\end{equation}
The guiding-center oscillator labels states within a Landau level, whereas the cyclotron oscillator changes the Landau-level index.
In this representation, multiplication by the coordinates and the leading Landau operator are
\begin{equation}
 Z=\mathfrak a^\dagger+\mathfrak b,
 \qquad
 \overline Z=\mathfrak b^\dagger+\mathfrak a,
 \qquad
 \mathcal L=4\mathfrak b^\dagger\mathfrak b.
 \label{eq:BFcoord}
\end{equation}
Here $\mathcal L$ is the local limit after the dimensionless Landau operator is divided by $N$.
Write $|a,\ell\rangle$ for a state with guiding-center occupation $a$ and cyclotron occupation $\ell$.
The LLL is the cyclotron vacuum, $\mathfrak b|a,0\rangle=0$, and $P_0$ denotes projection onto that vacuum for all guiding-center states.
The symbol $\mathfrak a$ is an operator, while the ordinary letter $a$ is an integer state label.
For the fused polynomial $Z^r$, the defect space is
\begin{equation}
 \mathcal C^{\rm BF}_r=
 \operatorname{span}\{|a,0\rangle:0\le a<r\}.
 \label{eq:BFdefect}
\end{equation}
The label $a$ is therefore also the fused defect-branch index.

Acting with the adjoint coupling on a fused defect state gives
\begin{equation}
 \overline Z^{\,r}|a,0\rangle
 =\sum_{s=0}^{a}\binom{r}{s}
 (\mathfrak b^\dagger)^{r-s}\mathfrak a^s|a,0\rangle.
 \label{eq:expand}
\end{equation}
The sum stops at $s=a$ because further guiding-center annihilation gives zero.
Each term has cyclotron occupation $r-s\geq1$, so different terms are orthogonal and the inverse $\mathcal L^{-1}$ contributes $[4(r-s)]^{-1}$.
Including the oscillator normalization factors gives
\begin{align}
 c_{r,a}
 &=\left\langle a,0\left|
 Z^r\mathcal L^{-1}(1-P_0)\overline Z^{\,r}
 \right|a,0\right\rangle
 \nonumber\\
 &=\frac{r!}{4}\sum_{s=0}^{a}
 \binom{r}{s}\binom{a}{s}\frac{1}{r-s},
 \qquad 0\le a<r.
 \label{eq:cra}
\end{align}
Rotational symmetry makes the matrix diagonal in the fused basis.
The coefficients are strictly ordered because, for $0\le a<r-1$,
\begin{equation}
 c_{r,a+1}-c_{r,a}
 =\frac{r!}{4}\sum_{s=1}^{a+1}
 \binom{r}{s}\binom{a}{s-1}\frac{1}{r-s}>0.
 \label{eq:strictordering}
\end{equation}

The coefficient $\beta_r$ uses the same dimensionless profile normalization as in the main text.
For a physical local coordinate $z$ near a zero on the sphere, write
\begin{equation}
 w(z)=\beta_r\left(\frac{z}{2R}\right)^r
 +\Ord\!\left(\left|\frac{z}{2R}\right|^{r+1}\right)
 \label{eq:betadef}
\end{equation}
in a regular local gauge.
Thus $\beta_r$ is the coefficient of the dimensionless local coordinate to order $r$; the coefficient of the dimensional monomial $z^r$ is $\beta_r/(2R)^r$.
For $w_{\rm f}=\mathcal N_{\rm f}v^r$, the leading relation $v\simeq z/(2R)$ in this gauge gives $\beta_r=\mathcal N_{\rm f}$.
Since $z/(2R)=N^{-1/2}Z[1+\Ord(N^{-1})]$, the local coupling has the leading form $\beta_rN^{-r/2}Z^r$.
The sphere-to-plane limit then gives
\begin{equation}
 D_N=|\beta_r|^2N^{-r-1}
 \operatorname{diag}_{0\le a<r}(c_{r,a})
 +\Ord(N^{-r-3/2}).
 \label{eq:DNasymptotic}
\end{equation}
The comparison is in operator norm after identifying the localized sphere states with the oscillator basis.
For the rotationally symmetric profile $v^r$, odd local corrections vanish and the remainder improves to $\Ord(N^{-r-2})$.
The expansion uses the sphere kernel in Eq.~\eqref{eq:Bergmankernel}, the dimensionless Landau energies in Eq.~\eqref{eq:LandauEnergy}, and the Wigner matrix elements in Sec.~\ref{sec:numerics}.

The normalization $(4\pi)^{-1}\int d\Omega\,|w_{\rm f}|^2=1$ gives $|\mathcal N_{\rm f}|^2=r+1$.
In the Haar-measure Wigner convention used for the numerical calculation, the same normalized profile is
\begin{equation}
 w_{\rm f}=\sqrt{r+1}\,D^{r/2}_{-r/2,r/2},
 \qquad |\beta_r|^2=r+1.
 \label{eq:betacoherent}
\end{equation}
The universal oscillator coefficients and the normalized crossing coefficients are related by
\begin{equation}
 \kappa_{r,a}=(r+1)c_{r,a}.
 \label{eq:kappadef}
\end{equation}
Table~\ref{tab:kappa} lists both quantities; the normalization factor $r+1$ is included in $\kappa_{r,a}$.

\begin{table}[t]
\caption{Universal fused-defect coefficients.}
\label{tab:kappa}
\centering
\begin{tabular}{c c c c}
\toprule
$r$ & $a$ & $c_{r,a}$ & $\kappa_{r,a}=(r+1)c_{r,a}$\\
\midrule
1 & 0 & $1/4$ & $1/2$\\
2 & 0 & $1/4$ & $3/4$\\
2 & 1 & $5/4$ & $15/4$\\
3 & 0 & $1/2$ & $2$\\
3 & 1 & $11/4$ & $11$\\
3 & 2 & $19/2$ & $38$\\
\bottomrule
\end{tabular}
\end{table}

\section{Full-multiband zero-energy crossings}
\label{sec:fullcrossing}

We now relate the eigenvalues of $D_N$ to zero-energy crossings of the full Hamiltonian in Eq.~\eqref{eq:HS}.
The distinction is important: a second-order energy shift alone does not establish the position or number of full-Hamiltonian crossings.

\subsection{Statement and proof strategy}

\paragraph{Theorem 1 (fused full-spectrum crossings).}
Fix the order $r\geq1$ and a nonzero fused profile $w=\mathcal N_{\rm f}v^r$.
Let $\lambda_0$ and $\Lambda$ be fixed positive bounds, independent of $N$, with $0<\lambda_0\leq|\lambda|\leq\Lambda$.
Assume that unmatched component-diagonal shifts have been removed at the $N^{-r-1}$ scale.
The statement concerns the resonantly matched Hamiltonian in Eq.~\eqref{eq:HS}, not that Hamiltonian with arbitrary additional diagonal disorder.
Let $0<\mu_{N,0}\leq\cdots\leq\mu_{N,r-1}$ be the eigenvalues of $D_N$.
There are positive constants $K,C,N_0$, independent of $N$ and uniform in the stated coupling interval, such that for $N\geq N_0$ there are exactly $r$ zero-energy-crossing detunings in $|\tau|\leq KN^{-r-1}$.
These detunings are negative and can be labeled so that
\begin{equation}
 \max_{0\le a<r}
 \left|\tau_{N,a}+\lambda^2\mu_{N,a}\right|
 \le C\lambda^2N^{-r-2}.
 \label{eq:crossingtheoremS}
\end{equation}
Each ordered eigenvalue of the reduced defect operator constructed below decreases strictly with $\tau$ in this window and crosses zero once.
The elimination steps preserve these crossings and their multiplicities in the full Hamiltonian.
For the normalized fused profile,
\begin{equation}
 -\frac{N^{r+1}}{\lambda^2}\tau_{N,a}
 \xrightarrow{N\to\infty}\kappa_{r,a}.
 \label{eq:crossinglimitS}
\end{equation}

The proof has two elimination steps, implemented by Schur complements, or Feshbach reductions~\cite{Feshbach,Grushin}.
We first eliminate all excited Landau levels and obtain an operator on the two LLLs.
We then eliminate the LLL states paired by $B_N$, leaving an $r$-dimensional operator on $\CN$.
The remainder is smaller than $D_N$ by one power of $N$, and a derivative bound fixes the number of zero crossings.
Constants denoted by $c$ or $C$ below may change between bounds; they remain independent of $N$ in the stated regime.

\subsection{Eliminating excited Landau levels}

Let $P_{N+r}^{(+)}$ project onto the upper LLL and define
\begin{equation}
 P=P_N^{(-)}\oplus P_{N+r}^{(+)},
 \qquad Q=1-P.
 \label{eq:PQ}
\end{equation}
Here $P$ keeps both LLLs, and $Q$ keeps all excited Landau levels in both components.
The projector $Q$ is distinct from the single-component Hamiltonians $Q_N^{(-)}$ and $Q_{N+r}^{(+)}$.
For $|\tau|\leq KN^{-r-1}$, the diagonal part of $QH_{N,r}(\tau)Q$ has a lower bound $cN$, while the Raman-coupling norm remains bounded.
The restricted inverse therefore satisfies, for sufficiently large $N$,
\begin{equation}
 \|(QH_{N,r}(\tau)Q)^{-1}\|\le CN^{-1}.
 \label{eq:Qinverse}
\end{equation}
All inverses of $QHQ$ act on $\Ran Q$, and $H$ abbreviates $H_{N,r}(\tau)$ in the following formulas.
Eliminating the $Q$ component from the zero-energy equation gives the Schur complement
\begin{equation}
 F_P(\tau)=PHP-PHQ(QHQ)^{-1}QHP.
 \label{eq:firstFeshbach}
\end{equation}
For example, the eliminated component is $Q\psi=-(QHQ)^{-1}QHP\psi$; inserting it into the $P$ equation gives $F_P(\tau)P\psi=0$.
Thus $F_P(\tau)$ and the full Hamiltonian have the same zero crossings, including their multiplicities.
Expanding the excited-level inverse around the uncoupled operator yields
\begin{equation}
 (QHQ)^{-1}=G_N^{(-)}\oplus G_{N+r}^{(+)}+\Ord(N^{-2}),
 \label{eq:horizontalinverse}
\end{equation}
where $G_{N+r}^{(+)}$ is the upper-component reduced resolvent, defined as in Eq.~\eqref{eq:Green-S}.
Because the holomorphic profile maps the lower LLL into the upper LLL, there is no direct lower-LLL coupling to upper excited levels:
\begin{equation}
 (1-P_{N+r}^{(+)})M_wP_N^{(-)}=0,
 \qquad
 P_N^{(-)}M_w^\dagger(1-P_{N+r}^{(+)})=0.
 \label{eq:holomorphicidentity}
\end{equation}
For $u\in\CN$, the reverse coupling to the lower LLL also vanishes, $P_N^{(-)}M_w^\dagger u=B_N^\dagger u=0$.
The leading correction on $\CN$ in Eq.~\eqref{eq:firstFeshbach} is consequently $-\lambda^2D_N$.

To control the error at the small crossing scale, one must retain the suppression at the dark-state endpoints of the virtual transition.
The sphere kernels and Wigner coefficients give
\begin{align}
 \|(1-P_N^{(-)})M_w^\dagger P_{\CN}\|&\le CN^{-r/2},
 \label{eq:defectmultbound}\\
 \|(G_N^{(-)})^{1/2}M_w^\dagger P_{\CN}\|&\le CN^{-(r+1)/2},
 \label{eq:halfgreenbound}\\
 \|P_{\CN}M_w(G_N^{(-)})^2M_w^\dagger P_{\CN}\|
 &\le CN^{-r-2}.
 \label{eq:twogreenbound}
\end{align}
In the fused basis, the Wigner selection rule makes the first lower cyclotron level accessible from branch $a$ equal to $\ell=r-a\geq1$.
The squared coupling matrix elements are finite products of factorial ratios, while each dimensionless inverse Landau energy is $[4\ell(N+\ell+1)]^{-1}=\Ord(N^{-1})$.
For fixed $r$, multiplication by the spin-$r/2$ profile connects only finitely many neighboring angular-momentum channels.
The finite sums then give the bounds above.
In particular, replacing the leading resolvent by the $\Ord(N^{-2})$ correction in Eq.~\eqref{eq:horizontalinverse}, with a defect coupling at both ends, changes the defect diagonal by $\Ord(N^{-r-2})$.

\subsection{Eliminating the paired LLL states}

Separate the upper LLL into states reached by $B_N$ and dark states:
\begin{equation}
 \LLL_{N+r}=\RN\oplus\CN,
 \qquad \RN=\Ran B_N.
 \label{eq:RNCN}
\end{equation}
The lower LLL and $\RN$ have equal dimensions and form the paired part of the projected coupling.
Write the polar decomposition as $B_N=U_NS_N$, where $S_N=(B_N^\dagger B_N)^{1/2}$ contains the coupling singular values and $U_N:\LLL_N\to\RN$ is unitary.
Using $U_N$ to identify these two spaces, the principal paired block becomes
\begin{equation}
 \mathcal P_N^{(0)}(\tau)=
 \begin{pmatrix}
 \tau & \lambda S_N\\
 \lambda S_N & -\tau
 \end{pmatrix}.
 \label{eq:pairedprincipal}
\end{equation}
The smallest singular value is of order $N^{-r/2}$, so an unweighted bound on this inverse would become large.
The relevant inverse is instead multiplied by a coupling factor $S_N$.
For a single singular value $s>0$,
\begin{equation}
 \left[\begin{pmatrix}\tau&\lambda s\\\lambda s&-\tau\end{pmatrix}\right]^{-1}
 =\frac{1}{\tau^2+\lambda^2s^2}
 \begin{pmatrix}\tau&\lambda s\\\lambda s&-\tau\end{pmatrix},
 \label{eq:scalarinverse}
\end{equation}
and
\begin{equation}
 \frac{|\lambda|s^2}{\tau^2+\lambda^2s^2}\le\frac1{|\lambda|},
 \qquad
 \frac{|\tau|s}{\tau^2+\lambda^2s^2}\le\frac1{2|\lambda|}.
 \label{eq:weightedinverse-scalar}
\end{equation}
Applying these inequalities in the eigenbasis of $S_N$ gives
\begin{equation}
 \left\|\mathcal P_N^{(0)}(\tau)^{-1}
 \binom{0}{S_Nv}\right\|
 +
 \left\|\mathcal P_N^{(0)}(\tau)^{-1}
 \binom{S_Nv}{0}\right\|
 \le C\|v\|,
 \label{eq:weightedinverse0}
\end{equation}
for every $v\in\LLL_N$.
The lower coupling bound $|\lambda|\geq\lambda_0$ makes $C$ uniform in the stated regime.
This is the weighted inverse estimate: the small singular value is accompanied by the coupling factor that cancels its potentially large inverse.

The excited-level elimination adds a correction $\Sigma_{\rm pair,N}$ to the paired block.
This is the paired self-energy: it records virtual excursions from paired LLL states into excited Landau levels and back.
On the sphere, dividing each LLL endpoint of Eq.~\eqref{eq:wignerME} by the corresponding singular factor in Eq.~\eqref{eq:exactsingular} leaves a Landau denominator and finite sums of normalized Clebsch--Gordan coefficients.
The resulting relative correction is of order $N^{-1}$ and has the factorized form
\begin{equation}
 \Sigma_{\rm pair,N}
 =\operatorname{diag}(S_N,S_N)\mathcal K_N
 \operatorname{diag}(S_N,S_N),
 \qquad \|\mathcal K_N\|\le CN^{-1}.
 \label{eq:relativepair}
\end{equation}
The operator $\mathcal K_N$ describes the remaining bounded correction and is distinct from the reproducing kernel $K_n(p,q)$.
Combining this factorization with Eq.~\eqref{eq:weightedinverse0} gives
\begin{equation}
 \|\Sigma_{\rm pair,N}
 (\mathcal P_N^{(0)}(\tau))^{-1}\|\le CN^{-1}.
 \label{eq:pairneumann}
\end{equation}
For large $N$, this bound permits a convergent Neumann expansion of the corrected inverse.
The complete paired block $\mathcal P_N=\mathcal P_N^{(0)}+\Sigma_{\rm pair,N}$ is therefore invertible and satisfies
\begin{equation}
 \left\|\mathcal P_N(\tau)^{-1}
 \binom{0}{S_Nv}\right\|
 +
 \left\|\mathcal P_N(\tau)^{-1}
 \binom{S_Nv}{0}\right\|
 \le C\|v\|.
 \label{eq:weightedinverse}
\end{equation}
This step does not require a separate comparison of $|\tau|$ with the smallest singular value.

The excited-level elimination also leaves a coupling $X_N:\CN\to\RN$ between dark states and the upper paired states.
Set $A_N=B_NB_N^\dagger|_{\RN}$, so that $A_N^{1/2}$ contains the range-side singular factors.
Since the direct LLL coupling vanishes on $\CN$, every contribution to $X_N$ contains a virtual Landau-level denominator.
Keeping the range-side coupling factor explicit gives
\begin{equation}
 X_N=A_N^{1/2}Y_N,
 \qquad
 \|Y_N\|\le CN^{-(r+2)/2}.
 \label{eq:XYbound}
\end{equation}
Here $Y_N$ is the remaining map from $\CN$ to $\RN$ after that factor is removed.
The estimate follows from the singular factors in Eq.~\eqref{eq:exactsingular} and the Wigner matrix elements in Eq.~\eqref{eq:wignerME}.
The defect endpoint contributes $N^{-r/2}$, the Landau inverse contributes $N^{-1}$, and the range singular factor is retained in $A_N^{1/2}$.
This form allows the weighted inverse bound to be used without introducing an uncontrolled inverse singular value.

\subsection{Reduced defect operator, error bounds, and crossing count}

Eliminating the paired block leaves the $r$-dimensional, detuning-dependent operator
\begin{equation}
 F_N^{\rm def}(\tau)
 =-\tau\one_{\CN}-\lambda^2D_N+R_N(\tau).
 \label{eq:defectpencil}
\end{equation}
Its zero eigenvalues locate the full-Hamiltonian crossings, while $R_N(\tau)$ collects the corrections beyond the leading second-order term.
The remainder $R_N$ is an operator and should not be confused with the range subspace $\RN$.
There are contributions from the correction to the excited-level resolvent, from returning through the paired block, and from further complementary resolvents.
Equations~\eqref{eq:twogreenbound}, \eqref{eq:weightedinverse}, and \eqref{eq:XYbound} give
\begin{equation}
 \|R_N(\tau)\|\le C\lambda^2N^{-r-2}.
 \label{eq:remainder1}
\end{equation}
This bound is one inverse power of $N$ smaller than the leading defect shift.
Differentiating an excited-level resolvent with respect to $\tau$ introduces one additional inverse.
For the paired block, one uses $\partial_\tau\mathcal P_N=\operatorname{diag}(\one,-\one)+\Ord(N^{-1})$ and the two weighted factors in Eq.~\eqref{eq:weightedinverse}.
Together these give
\begin{equation}
 \|\partial_\tau R_N(\tau)\|\le CN^{-1}.
 \label{eq:remainder2}
\end{equation}
The first bound controls the crossing positions; the derivative bound prevents repeated crossings of an ordered reduced branch in the stated window.

For $\tau_2>\tau_1$ and sufficiently large $N$,
\begin{equation}
 F_N^{\rm def}(\tau_2)
 \le F_N^{\rm def}(\tau_1)-\frac12(\tau_2-\tau_1)\one
 \label{eq:monotoneform}
\end{equation}
in quadratic-form order, meaning that the inequality holds in the expectation value of every vector in $\CN$.
Every ordered eigenvalue of $F_N^{\rm def}$ therefore decreases strictly with $\tau$.
At $\tau=0$, they are negative because $D_N>0$ and the remainder is smaller by one power of $N$.
The coefficient expansion gives the two-sided estimate
\begin{equation}
 cN^{-r-1}\le\mu_{N,a}\le CN^{-r-1}.
 \label{eq:Dtwosided}
\end{equation}
At $\tau=-KN^{-r-1}$, all ordered reduced eigenvalues are positive when $K$ is sufficiently large.
Each branch crosses zero once between these endpoints.
The two Schur complements preserve these crossings, including multiplicity, in the full Hamiltonian.
Finally, Weyl's eigenvalue bound, which limits an eigenvalue shift by the operator norm of the perturbation, applied to Eq.~\eqref{eq:defectpencil} gives Eq.~\eqref{eq:crossingtheoremS}.
Combining it with Eq.~\eqref{eq:DNasymptotic} gives Eq.~\eqref{eq:crossinglimitS}.

\section{Resolved unit-charge zeros and the $N^{-2}$ crossing law}
\label{sec:resolved}

The fused theorem concerns several units of winding at one point.
Here the same total winding is carried by distinct simple zeros.
We keep the fixed-coupling and resonance assumptions of Sec.~\ref{sec:fullcrossing}.

\paragraph{Theorem 2 (fixed resolved zeros).}
Let $w$ have $r$ distinct simple zeros $p_1,\ldots,p_r$ with a minimum pairwise separation bounded below by a positive constant independent of $N$.
Let $\delta z_j$ be the physical local displacement from $p_j$, rather than the position of the zero itself.
Using the dimensionless coefficient convention of Eq.~\eqref{eq:betadef}, write
\begin{equation}
 w(\delta z_j)=\beta_{1,j}\frac{\delta z_j}{2R}
 +\Ord\!\left(\left|\frac{\delta z_j}{2R}\right|^2\right),
 \qquad \beta_{1,j}\ne0.
 \label{eq:localsimple}
\end{equation}
The second-order effective operator $D_N$ then has $r$ eigenvalues
\begin{equation}
 \mu_{N,j}=\frac{|\beta_{1,j}|^2}{4}N^{-2}+\Ord(N^{-5/2}),
 \qquad j=1,\ldots,r,
 \label{eq:simpleD}
\end{equation}
up to relabeling of the branches.
The full Hamiltonian has exactly $r$ zero-energy crossings, counted with multiplicity, in a window of order $N^{-2}$ around resonance.
Their detunings obey
\begin{equation}
 \tau_{N,j}=-\frac{\lambda^2|\beta_{1,j}|^2}{4}N^{-2}
 +\Ord(N^{-5/2}).
 \label{eq:simpleroots}
\end{equation}
Here the local branch is labeled by its zero position $j$; in the large-$N$ limit, each branch has logarithmic exponent $2$.

\paragraph{Proof.}
Equation~\eqref{eq:simpleframe} gives an exact coherent-state basis of $\CN$.
Its normalized overlap, or Gram, matrix has elements $G_{jk}=\langle\widehat K_{p_j}^{(N+r)},\widehat K_{p_k}^{(N+r)}\rangle$ and obeys
\begin{equation}
 G_{jk}=\delta_{jk}+\Ord(e^{-cN}),
 \label{eq:simpleGram}
\end{equation}
where $\delta_{jk}$ is the Kronecker delta and $c>0$ is set by the minimum separation.
The matrix $G_{jk}$ is distinct from the reduced Landau resolvent $G_N^{(-)}$.
Orthonormalization therefore changes these coherent states only by exponentially small terms.

Near $p_j$, set $\delta z_j=\sqrt2\ell_B Z$.
Then
\begin{equation}
 w=\beta_{1,j}\frac{\sqrt2\ell_B}{2R}Z
 +\Ord\!\left(\frac{\ell_B^2}{R^2}|Z|^2\right)
 =\beta_{1,j}N^{-1/2}Z+\Ord(N^{-1})
 \label{eq:simplerescaling}
\end{equation}
for bounded $Z$.
The local defect state is the oscillator vacuum $|0,0\rangle$.
Equation~\eqref{eq:cra} with $r=1$ and $a=0$ gives $c_{1,0}=1/4$.
Two coupling factors supply $N^{-1}$, and the reduced Landau resolvent supplies another $N^{-1}$.
The higher local terms and curvature corrections are at most of relative order $N^{-1/2}$, while different fixed zero neighborhoods become exponentially decoupled in the localized description.
This gives Eq.~\eqref{eq:simpleD}.

The smallest nonzero singular values of $B_N$ now scale as $N^{-1/2}$, with a local sequence associated with each simple zero.
Repeating the weighted paired-block reduction with local vanishing order one gives
\begin{equation}
 F_{N,\rm split}^{\rm def}(\tau)
 =-\tau\one-\lambda^2D_N+\Ord(N^{-3}),
 \qquad
 \partial_\tau R_{N,\rm split}=\Ord(N^{-1}),
 \label{eq:splitpencil}
\end{equation}
where $R_{N,\rm split}(\tau)$ denotes the $\Ord(N^{-3})$ remainder in this split-profile reduction.
The same monotonicity argument gives one full-Hamiltonian zero crossing per reduced defect branch and establishes Eq.~\eqref{eq:simpleroots}. \hfill$\square$

The local order of the coupling zero fixes the endpoint exponent.
A fused order-$r$ zero gives $r$ jet branches with exponent $r+1$.
For $r$ separated unit-charge zeros, there is one local branch per zero and every branch has exponent $2$.

\section{Splitting crossover and the analytic two-zero model}
\label{sec:crossover}

\subsection{Leading crossover operator}

We vary the diameter $d$ of a small cluster near one pole while keeping its shape fixed.
This uniform dilation is the homothetic splitting used in the main text.
Write the physical local zero positions as
\begin{equation}
 z_j=d\,\zeta_j,
 \qquad j=1,\ldots,r,
 \label{eq:clusterzeros}
\end{equation}
where the dimensionless positions $\zeta_j$ are fixed and normalized by $\max_{j,k}|\zeta_j-\zeta_k|=1$.
For a nontrivial splitting cluster, $r\geq2$.
Under $z=\sqrt2\ell_B Z$, the leading local profile is represented by
\begin{equation}
 \begin{aligned}
 w_d(z)&=\beta_rN^{-r/2}q_x(Z),
 &q_x(Z)&=\prod_{j=1}^{r}(Z-x\zeta_j),\\
 x&=\frac{d}{\sqrt2\ell_B}
 =\frac{d\sqrt N}{2R}[1+\Ord(N^{-1})].
 \end{aligned}
 \label{eq:qx}
\end{equation}
Here $q_x$ is the monic polynomial describing the zero geometry, and $\beta_r$ carries the fixed leading profile normalization defined in Eq.~\eqref{eq:betadef}.
Equation~\eqref{eq:qx} specifies the leading rescaled profile, not an exact finite-$N$ replacement of the spherical coupling.
The omitted local curvature terms are included in the operator remainder below.

Write $q_x(Z)=\sum_{k=0}^{r}q_k(x)Z^k$.
Within the planar LLL, the adjoint of multiplication by $q_x$ acts as $\sum_k\overline{q_k(x)}\mathfrak a^k$.
The rescaled dark-state space is therefore
\begin{equation}
 \Cx=\ker\left[\sum_{k=0}^{r}
 \overline{q_k(x)}\mathfrak a^k\right],
 \qquad \dim\Cx=r.
 \label{eq:Cx}
\end{equation}
Let $P_x$ project onto $\Cx$, and recall that $P_0$ projects onto zero cyclotron occupation.
The factor $1-P_0$ therefore selects the excited cyclotron levels.
The rescaled second-order operator is
\begin{equation}
 \mathscr D_r(x)=P_xq_x(\mathfrak a^\dagger+\mathfrak b)
 [4\mathfrak b^\dagger\mathfrak b]^{-1}(1-P_0)
 q_x(\mathfrak a^\dagger+\mathfrak b)^\dagger P_x.
 \label{eq:Dcross}
\end{equation}
The inverse acts only at positive cyclotron occupation, just as the reduced resolvent in Eq.~\eqref{eq:Green-S} excludes the LLL.

For $x$ in a fixed bounded interval, the rescaled sphere kernels converge uniformly to the Bargmann--Fock kernels.
Using derivative states when zeros merge keeps the projector $P_x$ smooth through the fused limit.
Under this identification,
\begin{equation}
 D_N(d_N)=|\beta_r|^2N^{-r-1}\mathscr D_r(x)
 +\Ord(N^{-r-3/2}),
 \qquad
 d_N=\frac{2xR}{\sqrt N}[1+\Ord(N^{-1})].
 \label{eq:crossoverlimit}
\end{equation}
The limit holds at fixed rescaled separation $x$, so the corresponding physical separation $d_N$ decreases with $N$.
The defect-endpoint bounds, weighted paired-block bounds, and two-sided defect bound of Sec.~\ref{sec:fullcrossing} are uniform on the same fixed intervals through this local contraction and the confluent-frame construction.
If $f_a(x)$, $a=0,\ldots,r-1$, are the eigenvalues of $\mathscr D_r(x)$, the same reduction gives
\begin{equation}
 \tau_{N,a}(d_N)
 =-\lambda^2|\beta_r|^2N^{-r-1}f_a(x)+o(N^{-r-1}).
 \label{eq:crossoverroots}
\end{equation}
The crossover function in the main text includes the profile normalization: $\mathcal F_a(x)=|\beta_r|^2f_a(x)$.
For an exactly fused profile, $d=x=0$ for every $N$, so $q_x(Z)=Z^r$ and $f_a(0)=c_{r,a}$ give the fused coefficients.
These are leading crossover formulas; the uniform statement above does not claim control of every $x$ increasing without bound with $N$.

For large $x$, the local crossover states separate around the simple zeros $Z=x\zeta_j$.
The local slope of the polynomial at each zero is
\begin{equation}
 |q_x'(x\zeta_j)|^2
 =x^{2(r-1)}\prod_{k\ne j}|\zeta_j-\zeta_k|^2.
 \label{eq:qprime}
\end{equation}
Using the simple-zero oscillator coefficient $1/4$ gives
\begin{equation}
 f_j(x)=\frac14x^{2(r-1)}
 \prod_{k\ne j}|\zeta_j-\zeta_k|^2[1+o(1)].
 \label{eq:largex}
\end{equation}
Here $j$ labels the localized zero in the resolved limit.
At fixed physical $d>0$, fixed $R$, and fixed profile normalization, $x\propto\sqrt N\to\infty$, so the factor $N^{-r-1}x^{2(r-1)}$ has the $N^{-2}$ scaling of the resolved endpoint.
This matches the independent fixed-separated-zero result in Sec.~\ref{sec:resolved}.

A knee specifies where the logarithmic slope changes between its limiting regimes.
If a chosen knee criterion gives $x_{*,a}=\Ord(1)$ for branch $a$, the associated flux is
\begin{equation}
 N_{*,a}=4x_{*,a}^2\left(\frac{R}{d}\right)^2
 [1+\Ord(N_{*,a}^{-1})].
 \label{eq:kneegeneral}
\end{equation}
The scaling $N_*\sim(R/d)^2$ is common to these choices, while the prefactor depends on the branch, the cluster shape, and the chosen knee criterion.

\subsection{Analytic solution for two symmetric zeros}

For $r=2$, put the zeros at $z=\pm d/2$, so that $\zeta_1=1/2$ and $\zeta_2=-1/2$.
Then
\begin{equation}
 q_x(Z)=Z^2-\frac{x^2}{4},
 \qquad \sigma=\frac{x^2}{4},
 \label{eq:q2}
\end{equation}
where $\sigma$ is a dimensionless measure of the squared separation.
The rescaled dark-state condition is
\begin{equation}
 (\mathfrak a^2-\sigma)u=0.
 \label{eq:defecteq2}
\end{equation}
Let $|\alpha\rangle$ be a guiding-center coherent state, defined by $\mathfrak a|\alpha\rangle=\alpha|\alpha\rangle$, with real $\alpha=\sqrt\sigma=x/2$.
For $x>0$, normalized even and odd combinations are
\begin{align}
 |C_+\rangle&=
 \frac{|\alpha\rangle+|-\alpha\rangle}
 {\sqrt{2(1+e^{-2\sigma})}},
 \\
 |C_-\rangle&=
 \frac{|\alpha\rangle-|-\alpha\rangle}
 {\sqrt{2(1-e^{-2\sigma})}}.
 \label{eq:catstates}
\end{align}
The labels $+$ and $-$ in this subsection denote spatial parity, not the two internal components of Eq.~\eqref{eq:HS}.
The odd expression at $x=0$ is understood by continuity.
The even and odd states tend to the fused $a=0$ and $a=1$ states, respectively.

The sphere provides an exact geometric relation for the overlap of the two coherent states.
For upper-LLL states centered at geodesic distance $d$,
\begin{equation}
 \left|\left\langle\widehat K_{p_1}^{(N+2)},
 \widehat K_{p_2}^{(N+2)}\right\rangle\right|
 =\cos^{N+2}\!\left(\frac{d}{2R}\right).
 \label{eq:kerneloverlap2}
\end{equation}
Define $\sigma_N$ by matching this overlap to $e^{-2\sigma_N}$:
\begin{equation}
 \sigma_N=-\frac{N+2}{2}\log\cos\left(\frac{d}{2R}\right)
 =\frac{(N+2)d^2}{16R^2}
 +\Ord\!\left(\frac{(N+2)d^4}{R^4}\right).
 \label{eq:sigmaexact}
\end{equation}
The overlap identity is exact, whereas $\sigma=x^2/4$ is the leading local variable used in the crossover operator.
For a small cluster in the magnetic-length scaling limit, $\sigma_N$ approaches $\sigma$.
Matching overlaps accounts for this geometric relation but does not determine all finite-$N$ corrections to the effective operator or its crossings.

For a state $u\in\Cx$ with the cyclotron vacuum understood, Eq.~\eqref{eq:defecteq2} gives
\begin{equation}
 [(\mathfrak b^\dagger+\mathfrak a)^2-\sigma]u
 =(\mathfrak b^\dagger)^2u+2\mathfrak b^\dagger\mathfrak a u.
 \label{eq:q2action}
\end{equation}
The two terms have cyclotron occupations two and one.
Their inverse Landau denominators give the quadratic form
\begin{equation}
 \langle u|\mathscr D_2(x)|u\rangle
 =\frac14\|u\|^2+\|\mathfrak a u\|^2.
 \label{eq:q2quadratic}
\end{equation}
Evaluating this form in the even and odd states gives
\begin{equation}
 f_+(x)=\frac14+\sigma\tanh\sigma,
 \qquad
 f_-(x)=\frac14+\sigma\coth\sigma.
 \label{eq:fpm}
\end{equation}
These functions are exact eigenvalues of the leading local operator $\mathscr D_2(x)$.
They are not exact finite-$N$ eigenvalues of the spherical Hamiltonian.
As $x\to0$, $(f_+,f_-)\to(1/4,5/4)$, reproducing the fused coefficients.
As $x\to\infty$, both approach $x^2/4+1/4$, consistent with the resolved endpoint.

At fixed physical separation $d$, fixed $R$, and fixed coupling normalization, the leading variable obeys $\sigma\propto N$.
We use $s_\pm$, as in the main text, for the running logarithmic exponents:
\begin{equation}
 s_\pm(N)
 =-\frac{d\log|\tau_{N,\pm}|}{d\log N}
 =3-\frac{\sigma f_\pm'(\sigma)}{f_\pm(\sigma)}.
 \label{eq:runningexact}
\end{equation}
In this equation the functions in Eq.~\eqref{eq:fpm} are regarded as functions of $\sigma$, and the prime denotes differentiation with respect to $\sigma$, not $x$.
Both exponents tend to $3$ as $x\to0$ and to $2$ as $x\to\infty$, corresponding to the fused and resolved limits, respectively.
The crossover need not be monotonic: the even branch falls below $2$ before approaching the resolved limit, as shown in the main-text crossover figure.
Using $s_\pm=5/2$ to define the branch-dependent knee gives
\begin{equation}
 \sigma_{*,+}=0.29865,
 \qquad
 \sigma_{*,-}=1.246,
 \label{eq:sigmastar}
\end{equation}
and
\begin{equation}
 x_{*,+}=1.093,
 \qquad
 x_{*,-}=2.233,
 \qquad
 \frac{d_{*,+}}{\ell_B}=1.546,
 \qquad
 \frac{d_{*,-}}{\ell_B}=3.158.
 \label{eq:kneelB}
\end{equation}
Here $d_{*,\pm}$ is the separation assigned to the chosen knee at a given magnetic length, and the displayed numerical values are rounded.

An overlap-based spherical calibration is obtained by inserting the leading knee values into Eq.~\eqref{eq:sigmaexact} and solving for the distance:
\begin{equation}
 d_{*,\pm}=2R\arccos\!\left[
 \exp\left(-\frac{2\sigma_{*,\pm}}{N+2}\right)\right].
 \label{eq:kneesphere}
\end{equation}
This formula is the exact inversion of the overlap relation at the assigned $\sigma_{*,\pm}$.
It is not a claim that the full finite-$N$ spectral knee is known exactly, since the knee values came from the leading local crossover theory.
A finite-$N$ spectral calibration would also require the corresponding corrections to the crossing detunings.
Other knee definitions change the order-one coefficients, not the leading $N_*\sim(R/d)^2$ scaling.

\section{Spectroscopic protocol, state-resolved imaging, and physical units}
\label{sec:inverse}

For each flux $N$, sweep the detuning and determine where each defect branch crosses the common zero-energy reference.
Repeating the scan at fixed $R$, fixed zero positions, and fixed coupling normalization gives the logarithmic slope
\begin{equation}
 s_a(N)=-\frac{d\log|\tau_{N,a}|}{d\log N}.
 \label{eq:slopedef}
\end{equation}
The derivative describes the smooth large-$N$ dependence; for integer flux values, it is estimated from measurements at several $N$.
The endpoint laws give
\begin{equation}
 s_a\xrightarrow{N\to\infty}
 \begin{cases}
 r+1,&d=0\quad\text{(one exactly fused order-$r$ zero)},\\
 2,&d>0\ \text{fixed, with distinct unit-charge zeros}.
 \end{cases}
 \label{eq:slopeendpoint}
\end{equation}
Changing a fixed $\lambda$ rescales the leading crossing detunings by $\lambda^2$ but does not change these exponents.
For a partially split cluster, the knee estimates $d/R$ through Eq.~\eqref{eq:kneegeneral}, or through the overlap-based two-zero relation in Eq.~\eqref{eq:kneesphere} within its stated accuracy.
The leading crossover and the fixed-separated-zero endpoint have different limiting procedures, as described in Sec.~\ref{sec:crossover}.

The defect projector also describes the spatial distribution of the dark states.
For any orthonormal basis $\{\psi_a\}_{a=0}^{r-1}$ of $\CN$, define
\begin{equation}
 \rho_N^{\rm def}(p)=P_{\CN}(p,p)=\sum_{a=0}^{r-1}|\psi_a(p)|^2.
 \label{eq:defectdensity}
\end{equation}
Here $P_{\CN}(p,p)$ is the diagonal of the projector's position-space kernel; the sum is independent of the chosen orthonormal basis.
For fixed separated simple zeros, Eqs.~\eqref{eq:simpleframe} and \eqref{eq:simpleGram} imply that this density approaches a sum of coherent-state peaks centered at the zeros, each of width $\Ord(\ell_B)$.
The large-$N$ qualifier is needed because the coherent states are not exactly orthogonal at finite separation.
For a fused zero, the derivative-state basis instead gives $r$ internal profiles centered at the same point.
State-resolved imaging can locate these profiles when its spatial resolution is sufficient.
The spectral method compares the zero separation with the tunable magnetic length rather than with the imaging resolution.

To restore physical units, multiply the dimensionless crossing detuning by $E_0$.
Equations~\eqref{eq:gap} and \eqref{eq:crossinglimitS} give
\begin{equation}
 \frac{E_0|\tau_{N,a}|}{\Delta_{\rm LL}}
 \simeq\frac{\lambda^2\kappa_{r,a}}{4N^{r+2}}.
 \label{eq:unitratioS}
\end{equation}
The physical frequency shift is $E_0|\tau_{N,a}|/h$, where $h$ is Planck's constant.
This convention is the one used for the cold-atom and photonic estimates in the main text.
A fit at moderate $N$ retains a larger absolute signal but must account for the finite-$N$ deviations illustrated by the numerical results below.
Using a logarithmic slope removes a fixed multiplicative normalization, but does not remove the need to resolve the crossing positions or calibrate a detuning background.

\section{Scope beyond the sphere}
\label{sec:generalbase}

The sphere supplies the explicit formulas and is the geometry used for the theorems and numerical calculations above.
The endpoint exponent is determined locally, but an extension to another geometry requires assumptions about its low-energy states and gap.
On a compact magnetic Riemann surface, the assumptions used here are the following.
The magnetic line bundle is positive, so that the large-flux holomorphic ground space has the required localization.
The renormalized magnetic Laplacian has a gap $cN$ above this ground space.
After rescaling by the local magnetic length, its Bergman kernel and reduced Green kernel approach their Bargmann--Fock forms.
Finally, the coupling is a smooth holomorphic section whose first nonzero local derivative has order $r$.
The Bergman kernel describes projection onto the holomorphic ground space, while the reduced Green kernel describes inverse propagation outside it.
The cited magnetic-Laplacian and Bergman-kernel asymptotics provide the local inputs for this extension~\cite{Finski,MaMarinescu}.

Under these assumptions, the local packet width scales as $N^{-1/2}$.
An order-$r$ coupling zero supplies $N^{-r/2}$ at each of the two coupling vertices, and the reduced Green operator supplies $N^{-1}$.
The local second-order operator therefore scales as $N^{-r-1}$, with its coefficient determined by the corresponding normal-jet matrix in Eq.~\eqref{eq:cra} and the local profile normalization.
For another magnetic geometry, the local metric and field factors belong in that normalization; the sphere-specific coordinate factor $2R$ is not a universal length convention.
Fixed separated simple zeros give independent order-one neighborhoods and the $N^{-2}$ endpoint law.
The global index counts the total multiplicity, whereas the local coefficients distinguish its distribution among fused and separated zeros.
These are the assumptions behind the generality statement in the main text; no result for arbitrary gapless or nonholomorphic systems is asserted here.

For a possible planar realization, the relevant defect states must lie in the bulk, away from physical edges, with a well-resolved Landau gap and a Raman profile having the required local zero structure.
Raman-induced spin--orbital-angular-momentum coupling and vortex textures in spinor Bose--Einstein condensates~\cite{Chen2018} provide experimental ingredients, rather than an already realized test of these crossing laws.
This implementation prospect does not extend the compact-surface statement above to an arbitrary finite trapped condensate.

\section{Finite-Landau-level calculations and precision checks}
\label{sec:numerics}

\subsection{Hamiltonian matrix in the monopole-harmonic basis}

For component index $q=0,1$, the flux is $N+qr$.
At Landau level $\ell$, define the angular-momentum labels
\begin{equation}
 j_q=\frac{N+qr}{2}+\ell,
 \qquad \nu_q=\frac{N+qr}{2}.
 \label{eq:jq}
\end{equation}
We use $\nu$ for the right Wigner index so that it is not confused with the integer flux $n$ used in Eq.~\eqref{eq:LandauEnergy}.
For the normalized fused profile in Eq.~\eqref{eq:betacoherent}, the coupling matrix element is
\begin{align}
 &\langle j'm'\nu'|M_{w_{\rm f}}|jm\nu\rangle
 \nonumber\\
 &\quad=(-1)^{m'-\nu'}
 \sqrt{(2j'+1)(r+1)(2j+1)}
 \begin{pmatrix}
 j'&r/2&j\\-m'&-r/2&m
 \end{pmatrix}
 \begin{pmatrix}
 j'&r/2&j\\-\nu'&r/2&\nu
 \end{pmatrix}.
 \label{eq:wignerME}
\end{align}
The primed labels refer to the upper component and the unprimed labels to the lower component.
The two arrays are Wigner $3j$ symbols, obtained from the normalized Wigner triple-product formula~\cite{Varshalovich}.
The formula reproduces spinor-polynomial multiplication within the LLL.

Axial symmetry separates the fused Hamiltonian into independent blocks.
For defect branch $a=0,\ldots,r-1$, the magnetic labels are
\begin{equation}
 m_- =\frac N2+r-a,
 \qquad
 m_+=\frac{N+r}{2}-a.
 \label{eq:axialm}
\end{equation}
The target state in each block is the upper $\ell=0$ state.
The lower block begins at $\ell=r-a$, consistent with the local oscillator expansion in Eq.~\eqref{eq:expand}.

Numerically, we retain $0\leq\ell\leq\ell_{\max}$ wherever these levels occur in the block.
The diagonal entries are $E_{N,\ell}+\tau$ and $E_{N+r,\ell}-\tau$, using the dimensionless energies in Eq.~\eqref{eq:LandauEnergy}.
The off-diagonal entries are $\lambda$ times Eq.~\eqref{eq:wignerME}.
The angular-momentum triangle conditions restrict the coupling to nearby Landau indices, so the matrix is banded in $\ell$.
This finite-cutoff numerical matrix is distinct from the full Landau-tower operator used in the analytic crossing theorems.

\subsection{Crossing detunings and energy--detuning curves}

Some crossing detunings are of order $10^{-12}$, whereas other dimensionless matrix entries are of order $10^3$--$10^4$.
A direct eigenvalue calculation can lose accuracy when resolving the small eigenvalue against these much larger scales.
We therefore isolate the upper-LLL target state and write its axial block as
\begin{equation}
 H_a(\tau)=
 \begin{pmatrix}
 h_{00}(\tau)&h_{0R}(\tau)\\
 h_{R0}(\tau)&H_{RR}(\tau)
 \end{pmatrix}.
 \label{eq:axialblock}
\end{equation}
Here $h_{00}$ is the scalar target-state entry, $H_{RR}$ contains all other states in this finite block, and $h_{0R}$ and $h_{R0}$ couple the target to them.
The subscript $R$ here labels this numerical complement, not the paired-LLL subspace $\RN$.
Whenever $H_{RR}$ is invertible, zero is an eigenvalue precisely when the scalar residual vanishes:
\begin{equation}
 \mathcal R_a(\tau)=h_{00}(\tau)
 -h_{0R}(\tau)H_{RR}(\tau)^{-1}h_{R0}(\tau)=0.
 \label{eq:scalarresidual}
\end{equation}
This scalar $\mathcal R_a$ is distinct from the operator remainder $R_N(\tau)$ in Eq.~\eqref{eq:defectpencil}.
The residual is monotone in the crossing interval and is solved with a bracketed root finder.
Wigner symbols are evaluated in logarithmic-factorial form, and the complementary linear system is solved without explicitly forming its inverse.
We record the residual and, as a separate numerical check, the direct eigenvalue nearest zero.
A small residual alone is not a certification of an equal number of correct digits in the crossing detuning.

For the energy--detuning curves in the main text, the same scalar reduction is applied to $H_a(\tau)-E\one$ and solved for the eigenvalue $E$ at each detuning.
The plotted fused example uses $N=40$, $r=2$, $\lambda=0.30$, and $\ell_{\max}=18$.
Both $E$ and $\tau$ are dimensionless, and the displayed axes are rescaled by $N^3/\lambda^2$.
The energy-curve data and their zero-crossing checks are listed separately in Sec.~\ref{sec:data}; this calculation is not a finite-$N$ split-profile crossover simulation.

\subsection{Cutoff, coupling, and asymptotic checks}

Table~\ref{tab:cutoff} gives representative Landau-level cutoff tests for the scaled crossing detunings.
The displayed values are unchanged to the shown digits as $\ell_{\max}$ increases from $6$ to $18$.
These are finite-cutoff convergence checks, separate from the analytic large-$N$ limits.

\begin{table}[t]
\caption{Landau-level cutoff convergence for
$-N^{r+1}\tau_{N,a}/\lambda^2$ at $\lambda=0.30$.}
\label{tab:cutoff}
\centering
\begin{tabular}{c c c c c c}
\toprule
$N$ & $r$ & $a$ & $\ell_{\max}=6$ & $\ell_{\max}=12$ & $\ell_{\max}=18$\\
\midrule
160 & 1 & 0 & 0.484738319576 & 0.484738319576 & 0.484738319576\\
160 & 2 & 1 & 3.508172654675 & 3.508172654675 & 3.508172654675\\
160 & 3 & 2 & 33.71644380997 & 33.71644380997 & 33.71644380997\\
640 & 3 & 2 & 36.86611723527 & 36.86611723527 & 36.86611723527\\
\bottomrule
\end{tabular}
\end{table}

At $N=160$, changing $\lambda$ from $0.15$ to $0.30$ changes the scaled crossing detunings $-N^{r+1}\tau_{N,a}/\lambda^2$ by less than $4.0\times10^{-9}$ relatively over all fused branches.
This is consistent with the leading quadratic coupling law, as stated in the revised main text.

Table~\ref{tab:N640} lists all fused branches at $N=640$.
For $r=3$ and $a=0$, the crossing detuning is approximately $-1.04\times10^{-12}$ and the recorded scalar residual is below $3\times10^{-28}$.
For the same row, the distributed CSV records a direct eigenvalue nearest zero of approximately $-2.0\times10^{-28}$.
These recorded checks are small compared with the crossing detuning, but they do not certify $28$ decimal places or include every source of floating-point and truncation error.

\begin{table}[t]
\caption{All fused branches at $N=640$, $\lambda=0.30$, and
$\ell_{\max}=18$.  The scaled root is $-N^{r+1}\tau_{N,a}/\lambda^2$.}
\label{tab:N640}
\centering
\begin{tabular}{c c c c c}
\toprule
$r$ & $a$ & $\tau_{N,a}$ & scaled root & scaled root$/\kappa_{r,a}$\\
\midrule
1 & 0 & $-1.0901004\times10^{-7}$ & 0.4961168201 & 0.9922336401\\
2 & 0 & $-2.5272443\times10^{-10}$ & 0.7361132421 & 0.9814843228\\
2 & 1 & $-1.2659840\times10^{-9}$ & 3.6874457736 & 0.9833188729\\
3 & 0 & $-1.0367884\times10^{-12}$ & 1.9327136368 & 0.9663568184\\
3 & 1 & $-5.7120107\times10^{-12}$ & 10.6479596551 & 0.9679963323\\
3 & 2 & $-1.9776526\times10^{-11}$ & 36.8661172353 & 0.9701609799\\
\bottomrule
\end{tabular}
\end{table}

\section{Numerical data files}
\label{sec:data}

The accompanying data archive contains the following files.
All tabulated energies and crossing detunings use the dimensionless convention of Eq.~\eqref{eq:LandauEnergy}.
\begin{itemize}
 \item \texttt{fused\_crossings.csv}: crossing detunings, scaled values, asymptotic coefficients, ratios, scalar residuals, and direct eigenvalue checks for $N=20,40,80,160,320,640$ and all fused branches with $r=1,2,3$.
 \item \texttt{cutoff\_convergence.csv}: the Landau-level cutoff checks in Table~\ref{tab:cutoff}.
 \item \texttt{lambda\_scaling.csv}: the branch-resolved comparison between $\lambda=0.15$ and $0.30$ at $N=160$.
 \item \texttt{crossover\_r2\_exact.csv}: the analytic leading-local-model functions $f_\pm(x)$ and $s_\pm(x)$ from Eqs.~\eqref{eq:fpm} and \eqref{eq:runningexact}, sampled with $\Delta x=0.01$ over $0\leq x\leq8$.
 \item \texttt{energy\_detuning\_N40\_r2.csv}: the two fused $r=2$ energy branches at $N=40$, $\lambda=0.30$, and $\ell_{\max}=18$ used in the main-text energy--detuning plot.
 \item \texttt{energy\_detuning\_crossings.csv}: the zero crossings of those two branches and their comparison with the fused-crossing data.
 \item \texttt{energy\_cutoff\_checks.csv}: selected cutoff checks for the energy--detuning curves.
\end{itemize}
The existing crossover CSV retains the column names \texttt{gamma\_plus} and \texttt{gamma\_minus}; these columns contain $s_+$ and $s_-$, respectively.
Only the notation in the text has changed, not the stored numerical values or column names.
The word \texttt{exact} in the crossover filename refers to the analytic leading local functions, not to a finite-$N$ split-Hamiltonian calculation.

\footnotesize
\setlength{\bibsep}{0pt plus 0.2ex}